\documentclass{article}
\usepackage{quel-art}
\volnumber{20} \issuenumber{1} \edyear{2004}                             
\frompage{000} \topage{000}                                              
\recrevdate{1 January 2004}                                              
\def\rightmark{Entanglement from Solids or Neutrons}
\def\leftmark{M.O. Terra Cunha and V. Vedral}
 
\title{How to Extract Entanglement from a Piece of Solid or 
a Bunch of Neutrons} 
\authors{ 
{Marcelo O. Terra Cunha$^a$ and Vlatko Vedral %
\index{Terra Cunha, M.O.} 
\index{Vedral, V.} 
}\\[2.812mm]
{\normalsize
\hspace*{-8pt}The School of Physics and Astronomy, University of Leeds, Leeds LS2 9JT, UK\\[0.2ex] 
}}
 
\abstract{We review how to obtain spin entangled pairs of fermions from a Fermi gas.
          An experiment with neutrons is proposed in order to get such pairs.}

 
\makeindex

\newcommand{\de}[1]{\left( #1 \right)}

\newcommand{\DE}[1]{\left\{ #1 \right\}}

\newcommand{\eqref}[1]{(\ref{#1})}
\newcommand{\prl}{{\it Phys. Rev. Lett. }}
\newcommand{\pra}{{\it Phys. Rev. A }}
\newcommand{\prb}{{\it Phys. Rev. B }}
\newcommand{\rmp}{{\it Rev. Mod. Phys. }}

\newcommand{\ket}[1]{\left| #1 \right\rangle}
\newcommand{\bra}[1]{\left\langle #1 \right|}
\newcommand{\mean}[1]{\left\langle #1 \right\rangle}
\newcommand{\op}[1]{{\mathbf{#1}}}
\newcommand{\meanop}[1]{\mean{\op{#1}}}
\newcommand{\tr}{\mathrm{Tr}}
\newcommand{\sig}{\op{\sigma}}
\newcommand{\s}{\op{S}}

\newcommand{\eg}{{\it{e.g.}}, }
\newcommand{\etal}{{\it{et al.}}, }

\begin{document}
 
\maketitle

\section{Introduction}\label{intro}
Entanglement is a central aspect of modern discussions of quantum mechanics. Recognized in the early days\cite{EPR}, it was exclusively a matter of foundations for more than half a century\cite{Bell}. Its status changed with the birth of quantum information theory\cite{NC}, when it become a resource to be understood\cite{Und} and utilized in tasks like teleportation\cite{tele} as well as in some algorithms of quantum computation\cite{Shor,Grover}.

One natural task is to look for entanglement sources. The most usual strategies are to manipulate individual qubits and to control their interactions in order to generate entanglement, or to post select a special part of a quantum state, using our knowledge of some property to create entanglement in other degrees of freedom. Other tasks include the search for the consequences of entanglement for the properties of matter\cite{macro}.

In this article we will revise a proposed strategy to obtain spin entangled fermion pairs from a degenerated Fermi gas. We will address the natural difficulties in applying such ideas to electronic Fermi gases and propose the use of stored neutrons in order to experimentally investigate entanglement in Fermi gases.

\section{Pair entanglement in Fermi gases}\label{rev}
In this section we review two works, refs.~\cite{CEJP,Fuzzy}, which are the central contribution we presented in CEWQO. Details can be obtained in the cited references.
  
In Ref.~\cite{CEJP} the conditional state of a localized pair of fermions from a degenerated Fermi gas was studied. Consider $\ket{\Phi _0}$ as the ground state of the Fermi gas. In a second quantization notation, one has 
\begin{equation}
\ket{\Phi _0} = \prod _{s,p} \op{b}^{\dagger}_s\de{p} \ket{0},
\label{Phi0}
\end{equation} 
where $\op{b}^{\dagger}_s\de{p}$ is the creation operator of one fermion of spin $s$ and momentum $p$, and the product is over all modes with momentum smaller than the Fermi momentum $p
_f$. If one considers the two body density operator (up to normalization), conditioned to the detection of a pair of fermions at positions $r$ and $r'$, then\cite{CNYang}
\begin{equation}
\rho ^{\de{2}}_{ss',tt'} = \bra{\Phi_0}\op{\Psi}^{\dagger}_{t'}\de{r'}\op{\Psi}^{\dagger}_{t}\de{r}\op{\Psi}_{s'}\de{r'}\op{\Psi}_{s}\de{r}\ket{\Phi_0},
\label{rho2}
\end{equation}
where
\begin{equation}
\op{\Psi}_{s}\de{r} = \int _0^{p_f} \frac{d^3p}{\de{2\pi}^3} e^{ipr}\op{b}_s\de{p}
\label{Psir}
\end{equation}
is the annihilation operator localized at $r$.

The operator \eqref{rho2} can be considered as a two qubit density operator (see Appendix of Ref.~\cite{Fuzzy}) and application of Peres-Horodecki criterion\cite{PH} shows the existence of an {\emph{entanglement distance}}: if two fermions are extracted from the Fermi gas separated by less than this entanglement distance, then their spins are entangled. One physical point must be stressed: the position detection must be spin non-destructive for the scheme to make sense. One limiting case is very intuitive. If both fermions were detected at the same position, as they only have position and spin degrees of freedom, their spin state is necessarilly the singlet state $\ket{\uparrow \downarrow} - \ket{\downarrow \uparrow}$, which is maximally entangled. In Ref.~\cite{Fuzzy} it is shown how the increasing in the fermions separation corresponds to more mixing with a classically correlated state, which makes entanglement to vanish at and above this entanglement distance. Entanglement distance is essentially given by $k_{f}^{-1}$, where $k_f$ is the Fermi wavenumber.

One step forward was given in Ref.~\cite{Fuzzy} when the perfect localization condition was dropped out, and the notion of {\emph{fuzzy mesurements}} was introduced. Fuzzy measurements are characterized by a distribution $D\de{r-r''}$ and detection (annihilation) operators
\begin{equation}
\op{\Psi}_{s}\de{r} = \int \frac{d^3r''}{\de{2\pi}^3} \int \frac{d^3p}{\de{2\pi}^3} e^{ipr''} D\de{r-r''}\op{b}_s\de{p}.
\label{Psirfuzzy}
\end{equation}
Using this notion, entanglement was quantified and its dependence on the separation and on the broadening of gaussian detectors was studied (note that eq.~\eqref{Psir} corresponds to the situation when $D$ is a delta function). However, no pratical situation was devised.

It must be stressed that fuzzy measurements must be seen as {\emph{coherent}} detectors, in the sense that they have specific annihilation operators associated to them. They are not a simple bunch of incoherent localized detectors. Microscopic detectors tend to be closer to fuzzy measurements, while macroscopic detectors tend to add up incoherent signals.

\section{Electronic Fermi sea}\label{solids}
The simplest fermions one can think about are electrons and the simplest situation for them to be confined in a non-interacting Fermi gas is the case of conduction electrons in a metal. 

From a theoretic and abstract viewpoint, it can be done. The difficulty in proposing a real fuzzy measurement for such a gas comes from the entanglement distance they have, of the order of a few angstroms\cite{AM}. One should then be able to address a pair of conducting electrons as close as such a distance, without disturbing its spin state. This is not a simple direct task, and some other strategy needs to be devised in order to extract entanglement from the Pauli principle.

\section{A proposal: Neutrons}\label{neutrons}
For a confined Fermi gas, $k_f^3$ is proportional to the density of fermions\cite{BvK}. So, the root of the difficulties with conducting electrons is their high density. Neutrons can be used to create a Fermi gas suitable for the extraction of spin entangled pairs. We will describe such a proposal in this section.

The experiment we will propose keeps some similarities with an interesting recently performed measurement of evidences for neutron antibunching\cite{Pascazio}. In that experiment a beam of thermal neutrons is monochromatized by a cristal. The beam then incides in a beamsplitter and the coincidences on both outputs of the beamsplitter are recorded as a function of the optical path difference. A small deep is registered, which can be understood as a result of the fermionic antibunching (for each pair, the triplet spin component antibunches, while the singlet component bunches). The smallness of this deep can be interpreted as an indication that the vast majority of the recorded coincidences are accidental, in the sense that the two recorded neutrons were not really paired in the beam (in other words, they were independent), even when the optical paths agree. This is natural for a reasonably high flux of the order of $3000$ neutrons per second.

The one and very first difference in the experiment we propose is the previous creation of a confined Fermi gas. This is made to keep the situation the closest possible to the idealized Fermi gas, where boundary conditions imply the discreteness of the single particle levels. Ultra-cold neutrons\cite{UCN} are a good example of confined Fermi gas, but other alternatives are also possible.

Now we want a physical way of realizing the detection operator \eqref{Psirfuzzy}. Note that the $D$ distribution imply a wavepacket-like detector. This can be approximated by a sequence of a slit and collimators, where the emergent neutrons will have reasonably well defined position and momentum. The details of the slit and collimators will define the best model for $D$. A naive model is just a gaussian profile.

Naturally, neutrons will be emitted by this ``source'' in a statistical way. The flux can be estimated as $N r c e$, where $N$ is the number of trapped neutrons (decrescent with time), $r$ is the ratio between the area of the hole and the total area of the storing vessel, $c$ is the rate of colisions of neutrons with the walls, and $e\leq 1$ is an efficiency coefficient related to loses in the collimation process or in the detection (also effects of antibunching can be included in this factor). If one stores $N=10^5$ neutrons
in a bottle of one litre\cite{exp}, $c$ can be avaliatted as $c\approx 50~s^{-1}$. For a hole diameter of less than $1~mm$, one gets $r \approx 10^{-5}$. As $e$ depends on experimental details, one can say that roughly the flux will be of the order of few neutrons per second. The experiment intends to work with essentially simultaneously emitted neutrons. Such a concept depends on the definition of a simultaneity time window. This has to do with both, fundamental and practical aspects. From the first point, the broadening in momentum distribution of the emitted neutron defines a time scale which can be associated to its presence. From the other one, the response time and accuracy of the detectors play an important role. Suppose we allow a coincidence time of the order of $10^{-4}~s$. This value is large compared to the usual neutron detectors. From the values above, one can consider to have a mean value of the order of $10^{-3}$ neutrons per ``pulse'', which implies a coincidence rate of the order of $10^{-6}$ per counting or $10^{-2}$ coincidences per second. These numbers sugest the experiment to be feasible, despite difficult. The vessel should be fulled and detections made through a time like one hour (something around $30$ pairs of neutrons would be available). Then the preparation must be repeated and data recolected.

Up to this moment, we were not specific of which experiment should be done. If one just want to prove entanglement, something like a Bell experiment is enough. For completeness, we suggest to make a tomography of the spin state of neutron pairs\cite{time}. The procedure is detailed in the appendix. In any case, a beamsplitter should be used to separate the neutrons and spin polarizing detectors must be used in the equal optical path condition for recording coincidences.

\subsection{Other entanglement effects on neutrons}
We must emphasize that it is not the first time that one talks about entanglement on neutrons. There is a debate about some experiments on neutron scattering in which ``anomalous'' experimental cross-section  may\cite{Drey} or may not\cite{crit} be caused by short time entanglement among nucleons. Also violations of Bell inequalities were verified with neutrons\cite{BellNeut} which must be considered as entanglement on neutrons. However, while in the first case, the possibly entangled nucleons are subject to a strong environment which disentangles the system, and in the second experiment, the entanglement is between two degrees of freedom of the same neutron: momentum and spin; the experiment which we propose will be able to generate free flying spin entangled neutron pairs.

\section{Conclusions}\label{concl}
In this article we review how non-desctructive measurements can be used to extract entangled pairs of fermions from a Fermi gas. We discuss the difficulties in applying such ideas to conducting electrons in normal metals. We then suggest an experiment with neutrons capable of exhibit such a behaviour. It can be seen as a practical advance: a possible new source of entangled particles; and as a foundational interesting step, testing novel properties of fundamental constituents of matter.

\section*{Appendix:  Quantum state tomography}\label{app}
We want to describe a simple two-qubit tomographic procedure for neutron spins. We will first introduce a formalism in terms of expectation values of observables. Then we will convert it to neutron-countings since these are the available data.

The central idea can be understood for just one qubit. Denote by $\sig _0$ the $2\times 2$ identity matrix and by $\sig _j$, $j=1,2,3$, the usual Pauli matrices (by convention, latin indices run from 1 to 3, while greek indices include $0$). Note that $\DE{\sig _{\mu}}$ is a basis for the (real) vector space of ($2\times 2$) Hermitian  matrices. Moreover, $\sig _j$ are traceless matrices. So, any density operator can be written as
\begin{equation}
\rho = \frac{1}{2} \DE{\sig _0 + \sum _j b_j\sig _j},
\label{Bloch}
\end{equation}
in a unique way. The three real numbers $b_j$ form the so called {\emph{Bloch vector}}, and the positivity condition for $\rho$ implies that this vetor lies inside a unit sphere with respect to the usual Euclidean norm (the so called {\emph{Bloch sphere}}). Eq.~\eqref{Bloch} can be interpreted as a tomographic decomposition when one realizes that
\begin{equation}
b_j = \tr \DE{\sig _j \rho} = \mean{\sig _j},
\label{tomocoef}
\end{equation} 
which follows directly from the algebraic properties of Pauli matrices. This means that one only needs to measure three mean values to characterize a one qubit state. 

In terms of particle counting, it means that for each $j$ one needs to count two rates: $n_{j+}$ and $n_{j-}$ and the tomographic coefficient will be estimated by the ratio
\begin{equation}
b_j = \frac{n_{j+} - n_{j-}}{n_{j+} + n_{j-}}.
\label{tomocount}
\end{equation}
Whenever possible, the best way is to use polarizing beamsplitters (PBS) and two independent detectors, one in each output of the PBS. Measurements for different $j$ values involve the rotation of the PBS and the inclusion of quarter wave plates. If the PBS is not balanced, Eq.~\eqref{tomocount} must be properly adapted.

For generalizing this to two qubits, consider the matrices $\s_{\mu \nu} = \sig _{\mu}\otimes \sig_{\nu}$. Again, the set $\DE{\s _{\mu\nu}}$ is a basis for the (real) vector space of ($4\times 4$) Hermitian  matrices. In a more compact notation, one can write any density operator as
\begin{equation}
\rho = \frac{1}{4} \sum _{\mu,\nu} a_{\mu \nu}\s_{\mu \nu},
\label{tomo4}
\end{equation}
with the coefficients given by
\begin{equation}
a_{\mu \nu} = \tr \DE{\s _{\mu \nu}\rho} = \meanop{\s _{\mu \nu}},
\label{tomocoef4}
\end{equation}
which must now be interpreted. First of all, $\tr \DE{\s _{00}\rho} = 1$ and must not be obtained. The remaining $15$ coefficients can be divided into three sets: $a_{i0}$ refers only to the first neutron, $a_{0j}$ to the second, while $a_{ij}$ to correlations. Ideally, the first two sets could be measured without paying atention to coincidences. However, in the experiment we have in mind, we want to characterize the state of neutron pairs. So, it is necessary to use the second detector as a trigger for the relevant one.  

As in the one qubit case, the mean values must be translated into detection rates, now with only coincidences being registered. The coefficients can now be estimated as
\begin{eqnarray}
a_{ij} &=& \frac{n_{i+}n_{j+} - n_{i+}n_{j-} - n_{i-}n_{j+} + n_{i-}n_{j-}}{n_{i+}n_{j+} + n_{i+}n_{j-} + n_{i-}n_{j+} + n_{i-}n_{j-}}, \label{aij}\\
a_{i0} &=& \frac{n_{i+}n_{j+} + n_{i+}n_{j-} - n_{i-}n_{j+} - n_{i-}n_{j-}}{n_{i+}n_{j+} + n_{i+}n_{j-} + n_{i-}n_{j+} + n_{i-}n_{j-}}, \label{ai0}\\
a_{0j} &=& \frac{n_{i+}n_{j+} - n_{i+}n_{j-} + n_{i-}n_{j+} - n_{i-}n_{j-}}{n_{i+}n_{j+} + n_{i+}n_{j-} + n_{i-}n_{j+} + n_{i-}n_{j-}}, \label{a0j}
\end{eqnarray}
where each coefficient $a_{i0}$ and $a_{0j}$ will be estimated three times. Ideally these estimations should coincide. In practice they will coincide up to experimental errors and a statistical treatment can be used to refine them.

One important difficulty we want to address is that the use of PBS for two neutrons can give rise to difficulties in the registering of coincidences. In this case one will need to make independent measurements for each two complementar situations.

It is important to emphasize that, in a situation like the here proposed, where each outcome takes a relativelly long time, it can be interesting to consider better tomographic strategies, like the one proposed in ref.~\cite{Engl}.
 
\section*{Acknowledgment(s)}
The authors have the pleasure to thank the Organizers of CEWQO for invitation and reception. Discussions with other participants of the event contributed to the present work.
- MOTC recognizes support from CNPq, PRPq-UFMG, and the Millenium Institute for Quantum Information (CNPq-Brazil). VV thanks Engineering and Physical Sciences Research Council in UK for support. 
 
\section*{Notes} 
\begin{notes}
\item[a]
Permanent address: Departamento de Matem\'atica, Universidade Federal de Minas Gerais, Caixa Postal 702 Belo Horizonte 30123-970, Brazil;\\ 
E-mail: tcunha@mat.ufmg.br
\end{notes}

\vfill\eject

\begin{thebibliography}{99}
\bibitem{BvK}One can repeat the claculation shown, \eg ir ref.~\cite{AM} only changing the Born-von Karman periodic boundary conditions by the vanishing of the single particle wave funtion outside the gas recipient. The only effect is to add a factor $2$ in the calculation.
\bibitem{time}This would take roughly ten times more than a CHSH Bell inequality experiment, something that must be taken in account before performing data aquisition.
\end{thebibliography}
\end{document}